\documentstyle[12pt,preprint]{aastex}
\def\lineunits{ergs\ s$^{-1}$\,cm$^{-2}$}
\def\contunits{ergs\ s$^{-1}$\,cm$^{-2}$\,\AA$^{-1}$}
\def\Fvar{\ifmmode F_{\rm var} \else $F_{\rm var}$\fi}
\def\Rmax{\ifmmode R_{\rm max} \else $R_{\rm max}$\fi}
\def\tcent{\ifmmode \tau_{\rm cent} \else $\tau_{\rm cent}$\fi}
\def\tpeak{\ifmmode \tau_{\rm peak} \else $\tau_{\rm peak}$\fi}

\def\mm{\ifmmode \phantom{$-$} \else \phantom{$-$}\fi}

\def\kms{\ifmmode {\rm km\ s}^{-1} \else km s$^{-1}$\fi}
\def\Msun{\ifmmode M_{\odot} \else $M_{\odot}$\fi}
\def\Lsun{\ifmmode L_{\odot} \else $L_{\odot}$\fi}
\def\qo{\ifmmode q_{\rm o} \else $q_{\rm o}$\fi}
\def\Ho{\ifmmode H_{\rm o} \else $H_{\rm o}$\fi}
\def\ho{\ifmmode h_{\rm o} \else $h_{\rm o}$\fi}
\def\ltsim{\raisebox{-.5ex}{$\;\stackrel{<}{\sim}\;$}}

\def\vFWHM{\ifmmode v_{\mbox{\tiny FWHM}} \else
            $v_{\mbox{\tiny FWHM}}$\fi}
\def\CCF{\ifmmode F_{\it CCF} \else $F_{\it CCF}$\fi}
\def\ACF{\ifmmode F_{\it ACF} \else $F_{\it ACF}$\fi}
\def\Halpha{\ifmmode {\rm H}\alpha \else H$\alpha$\fi}
\def\Hbeta{\ifmmode {\rm H}\beta \else H$\beta$\fi}
\def\Hgamma{\ifmmode {\rm H}\gamma \else H$\gamma$\fi}
\def\Hdelta{\ifmmode {\rm H}\delta \else H$\delta$\fi}
\def\Lya{\ifmmode {\rm Ly}\alpha \else Ly$\alpha$\fi}
\def\Lyb{\ifmmode {\rm Ly}\beta \else Ly$\beta$\fi}

\def\ciii{\ifmmode {\rm C}\,{\sc iii} \else C\,{\sc iii}\fi}
\def\civ{\ifmmode {\rm C}\,{\sc iv} \else C\,{\sc iv}\fi}

\def\o5007{[O\,{\sc iii}]\,$\lambda5007$}

\begin{document}
\title{An Intrinsic Baldwin Effect in the 
\mbox{\boldmath H$\beta$}\ Broad Emission Line 
in the Spectrum of NGC 5548}

\author{Karoline M.\ Gilbert\altaffilmark{1} and Bradley M. Peterson}
\affil{Department of Astronomy, The Ohio State University,\\
    140 West 18th Avenue, Columbus, OH 43210-1173\\
Email:  kgilbert@astro.ucsc.edu, peterson@astronomy.ohio-state.edu}

\begin{abstract}
We investigate the possibility of an intrinsic Baldwin Effect (i.e., 
nonlinear emission-line response to continuum variations) in the broad 
\Hbeta\  
emission line of the active galaxy NGC 5548 using cross-correlation 
techniques to remove light travel-time effects from the data.  
We find a nonlinear relationship 
between the \Hbeta\ emission line and continuum fluxes that is in good 
agreement with theoretical predictions. We suggest that
similar analysis of multiple lines might provide a useful
diagnostic of physical conditions in the broad-line region. 
\end{abstract}
\altaffiltext{1}{Present address: University of California 
Observatories,
Lick Observatory, University of California at Santa Cruz, 
Santa Cruz, CA 95064.}

\setcounter{footnote}{0}

\section{Introduction}

Correlations between the continuum and emission-line properties
of active galactic nuclei (AGNs) afford potentially important
probes of the structure and physical conditions of the
broad-line region (BLR) and of the unobservable ionizing continuum
in these sources. Probably the best-established correlation
between continuum and emission-line properties is the ``Baldwin 
Effect''  (Baldwin 1977), an 
inverse, non-linear correlation between the equivalent width of the
broad UV emission lines, most notably \civ\,$\lambda1549$,
and the luminosity of the underlying 
continuum (e.g., Osmer, Porter, \& Green 1994).  
The Baldwin relationship can be parameterized as a simple power law 
of the form
\begin{equation}
\label{eq:basic}
F_{\rm line} \propto F_{\rm cont}^{\alpha},
\end{equation}
with index $\alpha \ltsim 1$.

Kinney, Rivolo, \& Koratkar (1990) and Pogge \& Peterson (1992)
investigated the effect of variability on this relationship
in \civ\ and \Lya. The results can be summarized as follows:
\begin{enumerate}
\item There is a {\em global} or object-to-object relationship
with slope $\alpha \approx 0.83$ for \civ\ and
$\alpha \approx 0.88$ for \Lya.
\item There is a rather flatter {\em intrinsic} relationship
with $\alpha \approx 0.4$ that describes how the line and
continuum relationship changes due to variability. At least
some  of the scatter in the global relationship is
attributable to variability and the flatter slope of
the intrinsic relationship. Moreover,
the scatter in this intrinsic relationship is largely 
attributable to light-travel time effects within the
BLR (i.e., changes in the line flux follow those in the
continuum flux with some measureable time delay).
\end{enumerate}

While the Baldwin Effect is well-established in
most of the strong UV lines, any global Baldwin Effect
in the Balmer lines seems to be quite weak
(e.g., Dietrich et al.\ 2002). However, to the
best of our knowledge, the question of 
whether or not there is an intrinsic relationship
in the Balmer lines has not been specifically addressed.
Our goal in this paper is to remedy this.
We adopt an approach similar to that of Pogge \& Peterson (1992) in 
attempting to remove light travel-time effects when determining the 
slope 
of the relationship.  
We will make use of numerous well-sampled optical spectra of NGC 5548
obtained by the International AGN Watch in this analysis
to show that an intrinsic Baldwin Effect is present in
\Hbeta.

\section{Data Analysis}
\subsection{The Data}

For this investigation,  we used ten years of spectra from the
International AGN Watch collaboration (Peterson et al.\ 2002 
and references therein).  These data are 
fairly inhomogeneous, with differences in 
spectral resolution, quality, and amounts of contaminating light from 
the host galaxy.  In order to minimize the effects of these  problems,
we restricted our analysis to the high-quality homogeneous data 
obtained with the Ohio State CCD spectrograph on the Perkins 1.8-m 
telescope at the Lowell Observatory in Flagstaff, Arizona, between 1988 
and  1998. The optical continuum and \Hbeta\ emission-line fluxes
for this subset are shown in Fig.\ 1.  
The continuum flux was measured in a 10\,\AA\ wide bin 
at a wavelength of 5100\,\AA\ in the rest 
frame of the galaxy, which has a redshift of $z=0.017$.  The continuum 
underneath the \Hbeta\ line was modeled by linear
interpolation between two windows on either side of the 
\Hbeta\ line, at wavelengths of 4710\,\AA\ 
and 5100\,\AA\ in the rest frame of the galaxy. The \Hbeta\ flux was 
then measured by integrating over the flux above this continuum.
The measurement uncertainties in these
spectral measurements were estimated by examining the
differences in fluxes measured on very short time scales
(i.e. less than or equal to 2 days). We assumed that 
flux differences on such short timescales are stochastic errors rather 
than real differences.  We estimated the fractional error in the 
integrated 
continuum and line fluxes to be 2.5\%, consistent with previous 
estimates of the error in these types of calculations (e.g., Peterson 
et al.\ 2002).  

\subsection{Light Travel-Time Effects}

In this investigation, it is important to  
take into account the effects of light-travel time.  
Continuum radiation emitted by the central continuum source 
(presumably an accretion disk)
is absorbed and reprocessed into emission-line 
radiation by diffuse BLR gas.  Due to light travel-time effects 
within the BLR, we observe a time delay of order days to weeks between 
variations in the continuum and the response of the emission lines. 
A simple linear model (Blandford \& McKee 1982) to describe the 
emission-line response to continuum variations can be written as 
\begin{equation}
\label{eq:transfer}
L(t) = \int \Psi(\tau) C(t-\tau)\, d\tau,
\end{equation}
where $C(t)$ and $L(t)$ are the continuum and emission-line
light curves, respectively,
and $\tau$ is the time delay, and
$\Psi(\tau)$ is the transfer function. The transfer function 
depends on the size, geometry, and structure of the BLR.  Under typical 
conditions, the centroid of the transfer function
is essentially the mean response time of an emission line to 
changes in the continuum.
The mean response time, or lag, for an emission line can be found by 
cross-correlation of the light curves.  

In the case of NGC 5548, multiple years of monitoring optical continuum
and emission-line variability has shown that the \Hbeta\ time
lag changes over time. More precisely, the lag varies with the
mean continuum level in a fashion that clearly suggests that
the response is dominated by gas in which physical conditions
are optimal for the \Hbeta\ line. Thus, the mean response time for
a given line is a dynamic quantity and eq.\ (\ref{eq:transfer}) is an overly
simplistic model of the line response. However, it seems to be 
sufficient to account for time-dependent changes in the lag by taking 
yearly averages (Peterson et al.\ 2002). 

We characterize the time delay by the value of the
centroid of the optical continuum/\Hbeta\ emission-line
cross-correlation functions for each year. The cross-correlation 
centroids are taken from Table 8 of Peterson et al.\ (2002);
values range from a low of 11.37 days in 1992 to a high
of 26.44 days in 1998.
For each point in the \Hbeta\ light curve, a corresponding 
lag-corrected continuum point was determined.  The lag was subtracted 
from the time of each \Hbeta\ measurement to identify the time at which 
the optical continuum was best-correlated with the \Hbeta\ 
emission-line flux.  The continuum flux at this time was determined by linear 
interpolation between the actual observations immediately preceding and 
following this time.  Errors on interpolated continuum points were 
estimated by adding the errors of the preceding and following continuum 
points in quadrature, which probably slightly overestimates the
actual uncertainty.

\subsection{Fitting Procedure and Results}

In Fig.\ 2, we show the total \Hbeta\ flux as a function of
optical continuum flux, corrected for time delay as described in the previous
section. The relationship is clearly non-linear. However, we
have not yet accounted for constant spectral contaminants 
that have to be removed from 
the observed optical continuum and emission-line fluxes.  
The two major contaminants to the spectrum of importance to our 
investigation are (a) the contribution of the host galaxy to the 
optical continuum and (b) the \Hbeta\ narrow-line contribution to the 
total \Hbeta\ flux. Assuming that these are the only two necessary 
corrections, we can write eq.\ (\ref{eq:basic}) in terms of the observables,\
namely the total continuum flux $F_{\lambda}(5100\,{\rm \AA})$
(i.e., AGN plus host galaxy) and the total \Hbeta\ flux
$F(\Hbeta_{\rm total})$ (i.e., broad component plus narrow component)
as
\begin{equation}
\label{eq:baldwin}
F(\Hbeta_{\rm broad}) =
F(\Hbeta_{\rm total}) - F(\Hbeta_{\rm narrow}) =
A(F_{\lambda}(5100\,{\rm \AA}) - F_{\rm galaxy})^{\alpha}.
\end{equation}

We can approach eq.\ (\ref{eq:baldwin}) 
in two ways, first by a complete non-linear
least squares treatment, and second by fixing the host galaxy and
narrow-component fluxes, which are well-constrained by independent
analyses, which reduces the problem to a linear least-squares
exercise. We consider both of these in turn.

\paragraph{\it Non-Linear Least Squares Fit.}
We employed a Marquardt algorithm as described by
Press et al.\ (1989) to determine the constant components
$F(\Hbeta_{\rm narrow})$ and $F_{\rm galaxy}$, the
scaling factor $A$, and the Baldwin Effect index
$\alpha$, all as defined in eq.\ (\ref{eq:baldwin}). 
The values of these parameters are 
shown in Table 1, along with uncertainties estimated by
bootstrap resampling. The best-fit locus is shown in Fig.\ 2.

\paragraph{\it Simple Power-Law Fit.} We made use of the fact that
two of the parameters in eq.\ (\ref{eq:baldwin}) 
can be accurately determined by
independent means. The host galaxy 
flux at 5100\,\AA\ is estimated to be 
$F_{\rm galaxy}=(3.37\pm0.54) \times 10^{-15}$\,\contunits\ 
(Romanishin et al.\ 1995) by using a model of the host-galaxy surface
brightness distribution obtained by careful deconvolution of
the nuclear point source and the host galaxy.

We used a value for the narrow \Hbeta\ line component of 
$F(\Hbeta_{\rm narrow}) = 6.7\times10^{-14}$\,\lineunits\, which was 
estimated by trial-and-error scaling of the 
[O\,{\sc iii}]\,$\lambda5007$ emission-line profile and subtracting 
the scaled profile at the position of the \Hbeta\ narrow line. This 
works best on high-quality data obtained during a low-flux state when 
the narrow component of \Hbeta\ is relatively prominent.
A plausible narrow-\Hbeta\ model was found by scaling the
[O\,{\sc iii}]\,$\lambda5007$
emission line by 0.12 ($\pm0.01$), 
as shown in Fig.\ 3. After subtraction of these
constant components, eq.\ (\ref{eq:baldwin}) 
reduces to a simple linear least squares problem
\begin{equation}
\label{eq:powerlaw}
\log F(\Hbeta_{\rm broad}) = \log A +
\alpha \log \left[ F_{\lambda}(5100\,{\rm \AA}) -
F_{\rm galaxy}\right].  
\end{equation}
The linear regression was performed using the BCES (bivariate 
correlated errors and 
intrinsic scatter) program described by Akritas \& Bershady (1996).  
This technique accounts for measurement errors in both parameters. 
The results for BCES(Y$\|$X) regression
are given in Table 1, again with uncertainties in the
fitted parameters determined by bootstrap resampling. The best
fit function is shown in Fig.\ 4.
 
\section{Discussion}
Both methods of fitting show a clear Baldwin Effect, with
$\alpha \approx 0.65$ for the simple linearized fit and
$\alpha \approx 0.53$ for the non-linear least-squares fit,
in both cases rather steeper than the values found by
Pogge \& Peterson (1992) for \civ\ and \Lya. 
The fit shown in Fig.\ 4 raises concern about either
the parameterization of the relationship or the
estimate of the background components (narrow-line
\Hbeta\ and host-galaxy); in other words, if the 
independent estimates of the background components are
correct, then as shown in Fig.\ 4, 
most of the low-flux points fall below the best-fit
line, indicating that the slope of the relationship
must be different in this regime. While the non-linear
least-squares fit (Fig.\ 2) does not have this problem
at the low-flux end, the  background estimates from
the fit are in both cases
much too high, although certainly the constraint
on the host-galaxy flux is less stringent than that
on the \Hbeta\ narrow component. To pursue this, if
we adjust upwards from the Romanishin et al.\ (1995)
estimate of the host-galaxy flux,
we find the best overall fit (as measured by 
$\chi^2$) is achieved for a value
of $4.6\times10^{-15}$\,\contunits
(yielding $\alpha \approx 0.52$), which is, not surprisingly,
close to the host-galaxy estimate we obtained 
(Table 1) by the non-linear least-squares
fit to eq.\ (\ref{eq:baldwin}), the only difference being whether
or not the narrow-line contribution is a free parameter or
held fixed. A galaxy contribution of 
$4.6\times10^{-15}$\,\contunits\ 
exceeds the Romanishin et al.\ (1995) estimate
by $\sim 2\sigma$, but it is still (although only 
slightly) below the minimum published optical continuum flux
(Peterson et al.\ 2002). However, during the 2002 observing
season, NGC 5548 went into the lowest-flux state recorded over
the last 30 years, reaching a minimum 5100\,\AA\ flux of 
$(3.94 \pm 0.10) \times10^{-15}$\,\contunits\ 
on JD2452471 (Sergeev 2002, private communication).
Given this result, we doubt that the estimate of
Romanishin et al.\ (1995) significantly underestimates
the host-galaxy contribution; it seems more likely that our
parameterization is oversimplified.

While Fig.\ 4 suggests that the relationship between 
the continuum and the \Hbeta\ emission line changes at low flux levels,
it is also possible that the relationship simply varies somewhat with 
time. As all of the low-flux points below the
best-fit lines are from a single epoch (the fourth year of this
monitoring program, during the period JD 2448713 -- 2448848),
we are unable to discriminate between flux dependence and time 
dependence. A complete analysis of the most recent faint-state
data should clarify this.

Finally, we as show in Fig.\ 3, our best estimate of
the \Hbeta\ narrow component, based on a profile decomposition,
is  $6.7\times10^{-14}$\,\lineunits. This is much lower
than the non-linear least-squares result of 
$1.8\times10^{-13}$\,\lineunits\ (Table 1; also see Fig.\ 2).
It is very unlikely that the \Hbeta\ narrow component 
based on profile decomposition is
underestimated. However, the non-linear least-squares result
may include other less-variable contaminants to the
broad \Hbeta\ profile, and this is under investigation.

Depending on how we account for the constant components,
we find possible power-law indices (eq.\ \ref{eq:baldwin}) in
the range $0.52 < \alpha < 0.65$.  In any 
case, we see clear evidence for an intrinsic Baldwin Effect in \Hbeta, 
i.e., the line flux does not increase linearly with the continuum flux.  
To explore the physical nature of this relationship, however, we need 
to relate the \Hbeta\ flux to the ionizing continuum flux rather than 
the optical continuum flux, since the former actually drives the 
emission-line variations.  While the ionizing flux 
($\lambda < 912$\,\AA) is not directly observable, 
measurements closer to the ionizing
continuum, at 1350\,\AA, are available 
from the {\it International Ultraviolet Explorer} archive.  
Assuming a relationship of the form
\begin{equation}
\label{eq:uvopt}
F_{\rm opt} \propto F_{\rm UV}^{\beta},
\end{equation}
Peterson et al.\ (2002) find $\beta \approx 0.56$, based
on contemporaneous pairs of 1350\,\AA\ ($F_{\rm UV}$)
and 5100\,\AA\ ($F_{\rm opt}$) continuum fluxes.  
However, by using
the BCES regression, we found a slightly 
steeper slope, $\beta = 0.67$, which we will use in this discussion. 
We also note that in the case of NGC~5548, 
Galactic extinction is a small enough effect that it 
makes no difference in the slope $\beta$ of eq.\ (\ref{eq:uvopt}).

Our analysis thus far has been based only on
optical data, and we have assumed specifically that
$F(\Hbeta) \propto F_{\rm opt}^{\alpha}$. Combining this
with eq.\ (\ref{eq:uvopt}). we can write the
relationship between the \Hbeta\ emission line flux and the UV 
ionizing continuum as
$F(\Hbeta) \propto F_{\rm UV}^{\alpha\beta }$, so that
$0.35 \ltsim \alpha\beta \ltsim 0.44$.

\section{The Effect of Geometric Dilution}

The emission-line light curve is a convolution of the
continuum light curve with a transfer function, as shown
in eq.\ (\ref{eq:transfer}). The effect of the transfer function is to
shift and broaden the emission-line light curve in time relative to
the continuum curve; the effect of making the BLR larger is
not only to shift the cross-correlation lag to larger values,
but also to smear out the response in time, reducing the
apparent amplitude of the response at any given time.
We therefore consider here how important an effect this
geometrical dilution of the line response might be
in this particular case. We attempt to estimate this by
convolving the real optical continuum light curve with a transfer function
for a thin spherical shell, since this is a rectangular function
that will have a relatively large dilution
of the line response through time-smearing effects,
and repeating the analysis described above.

The effect of geometric dilution on the amplitude of variability of 
emission lines was explored by convolving the NGC 5548 continuum with 
thin-shell transfer functions of radius between 1 and 10 light days.  
We then used the original continuum data and the simulated \Hbeta\ 
emission to calculate the slope of the Baldwin Effect as discussed above.  
We found that the slope does decrease slightly with increasing geometric 
dilution; however, it does so at a very small rate.  
The difference in $\alpha$ is found to be only about 0.05
for rectangular transfer functions of half width 1 and 10 days.
We tested transfer 
functions for other geometries, and found that, as expected, 
the effects of smearing were even
smaller for narrower profile transfer functions.

\section{Comparison with Theoretical Predictions}

We can compare our empirical results on the intrinsic Baldwin Effect
in \Hbeta\ with the theoretical 
predictions made in a series of photoionization calculations by Korista 
et al.\ (1997; see their Fig.\ 3g). 
The quantity $\Phi{\rm (H)}$ is defined as the number of 
ionizing photons arriving at the inner face of a BLR cloud
per square cm per sec. If the shape of the ionizing continuum
is fixed, the emission-line spectrum will be a function of
$\Phi{\rm (H)}$, which is proportional to the photoionization
rate, and of the particle density $n$ (cm$^{-3}$), which is
proportional to the recombination rate, as well as
chemical abundances and cloud column density. Korista et al.\
show the strength of various emission lines normalized by
the continnum flux (i.e., the line equivalent widths) for
various values of $\Phi{\rm (H)}$ and $n$ for different
column densities and standard assumptions about elemental
abundances. 
For NGC~5548, $\log \Phi{\rm (H)} =20$ at a distance of 12.6 light days from 
the central source (Korista \& Goad 2001).  
The equivalent width used by Korista et al.\ 
is defined as the ratio of the \Hbeta\ flux to the 
continuum flux at 1215\,\AA.  Using the above relationships, 
the equivalent width should be
\begin{equation}
W = \frac{F(\Hbeta)}{F_{\rm UV}} \propto 
F_{\rm UV}^{\alpha \beta - 1}.
\end{equation} 

Using our values of $\alpha$ and $\beta$, 
we find that we expect the equivalent
width to be proportional to the UV flux to the $-0.56$ or $-0.65$ 
power, depending how we account for the constant components.
In other words, an order of 
magnitude change in $\Phi{\rm (H)}$ should 
change the equivalent width by around $-0.60$ dex.  
From Fig.\ 3g of Korista et al.\ (1997), one can see that for 
$\log \Phi{\rm (H)} \approx 20$ and  
for hydrogen particle densities $n$(H)(cm$^{-3}$) in the range
$11 \ltsim \log n{\rm (H)} \ltsim 13$,
the slope of the $\log W / \log \Phi{\rm (H)}$ relationship
is consistent with our results.  
The grids\footnote{\raggedright
The photoionization grids calculated by Korista et al.\ 
(1997) can be found on the web at 
http://www.pa.uky.edu/$\sim$korista/grids/grids.html.}    
 were investigated to find the slope of the 
relationship for varying values of column density and continuum shape 
with solar metallicity.  Since the value of $\log \Phi{\rm (H)} 
\approx19.6$ 
at a radius of 20 light days, where typically the \Hbeta\ response
is maximized in  NGC 5548, 
we determined the slope of the grid for varying values of the 
hydrogen particle density by linear interpolation along the grid.
We found that regardless of the model, the
slope of the $\log W / \log \Phi{\rm (H)}$ relationship reaches a
maximum in the range $-0.57$ to $-0.60$ when the value of the 
hydrogen particle density is in the
range $12 \ltsim \log n{\rm (H)} \ltsim 12.5$.

\section{Conclusions}

We have investigated the relationship 
between the variable \Hbeta\ emission line and optical continuum of NGC 5548.  
We discovered that an intrinsic Baldwin Effect does exist;
after for correcting for light travel-time effects, we find that
the amplitude of broad \Hbeta\ variation is less than the
amplitude of variation of the underlying optical continuum.
By combining these results with
the relationship between the amplitudes of variability of the
optical and UV continuum, we determine the relationship between 
the \Hbeta\ emission-line flux and the UV continuum flux, 
and find that our result is in good agreement with the
predictions of photoionization calculations, even accounting
for smearing of the line response by light travel-time effects
within the \Hbeta-emitting region. 
We suggest that similar analysis of multiple lines have
potential as diagnostics of the physical nature of the BLR.
        
We are grateful for support of this work through NASA
grant NAG5-8397. We thank Sergey Sergeev of Crimean
Astrophysical Observatory for data on the recent faint
state of NGC 5548. We are grateful to an anonymous 
referee whose comments led to an improved presentation.



\clearpage

\begin{figure}
\epsscale{0.78}
\plotone{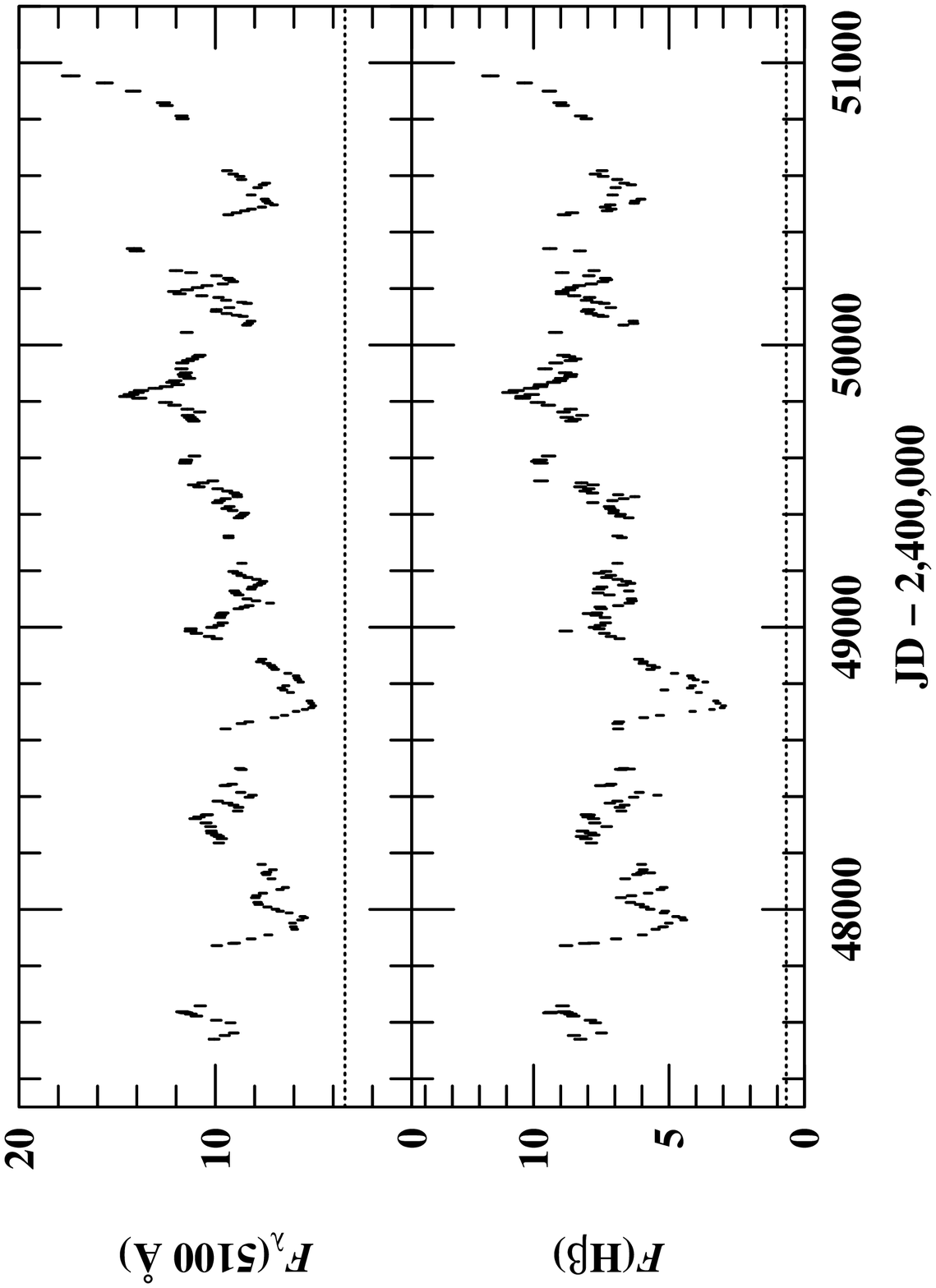}
\caption{The optical continuum (upper panel) and
\Hbeta\ (lower panel) light curves of NGC 5548 from 1989 December
to 1998 June. 
The continuum fluxes
are in units of $10^{-15}$\,\contunits, and the line fluxes
are in units of $10^{-13}$\,\lineunits. The dotted horizontal lines
show the independent estimates of the constant components,
specifically the continuum
contribution from the host galaxy starlight (Romanishin et al.\ 1995)
in the upper panel, and the narrow-line \Hbeta\ contribution
in the lower panel.
Flux measurements are in the observer's reference frame and are uncorrected
for Galactic extinction.}
\end{figure}

\clearpage

\begin{figure}
\epsscale{0.8}
\plotone{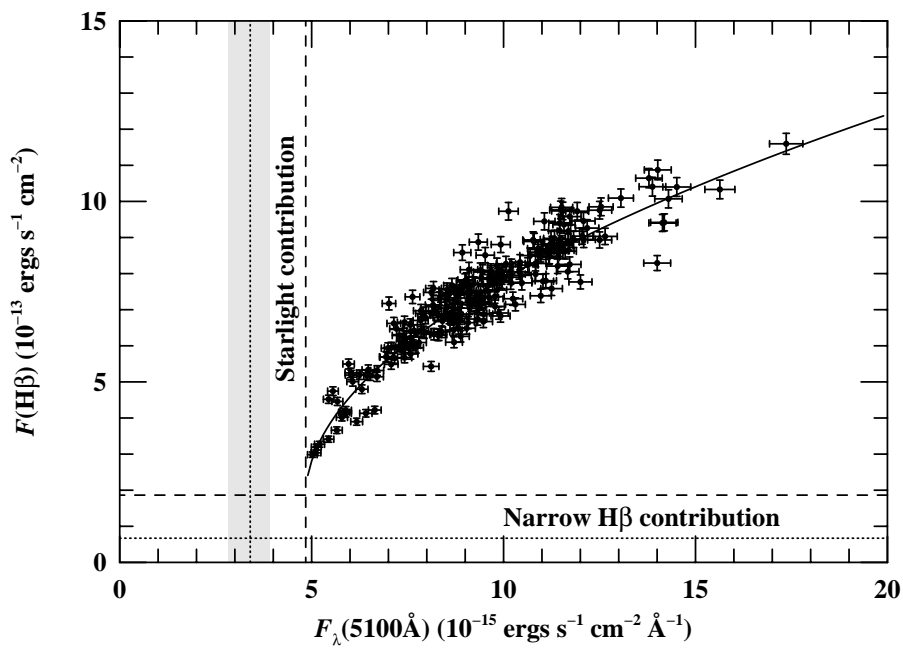}
\caption{\Hbeta\ vs.\ optical continuum flux for NGC 5548.
The solid line shows the best fit to these data using
a simple non-linear form (eq.\ \ref{eq:baldwin}). The dashed lines show
the constant components (host galaxy flux and \Hbeta\ narrow-line
contribution) based on this non-linear fit. The dotted lines
show the estimates of these quantities based on independent
estimates, with the shaded area showing the $1\sigma$
uncertainty in the host-galaxy flux.}
\end{figure}

\clearpage
\begin{figure}
\plotone{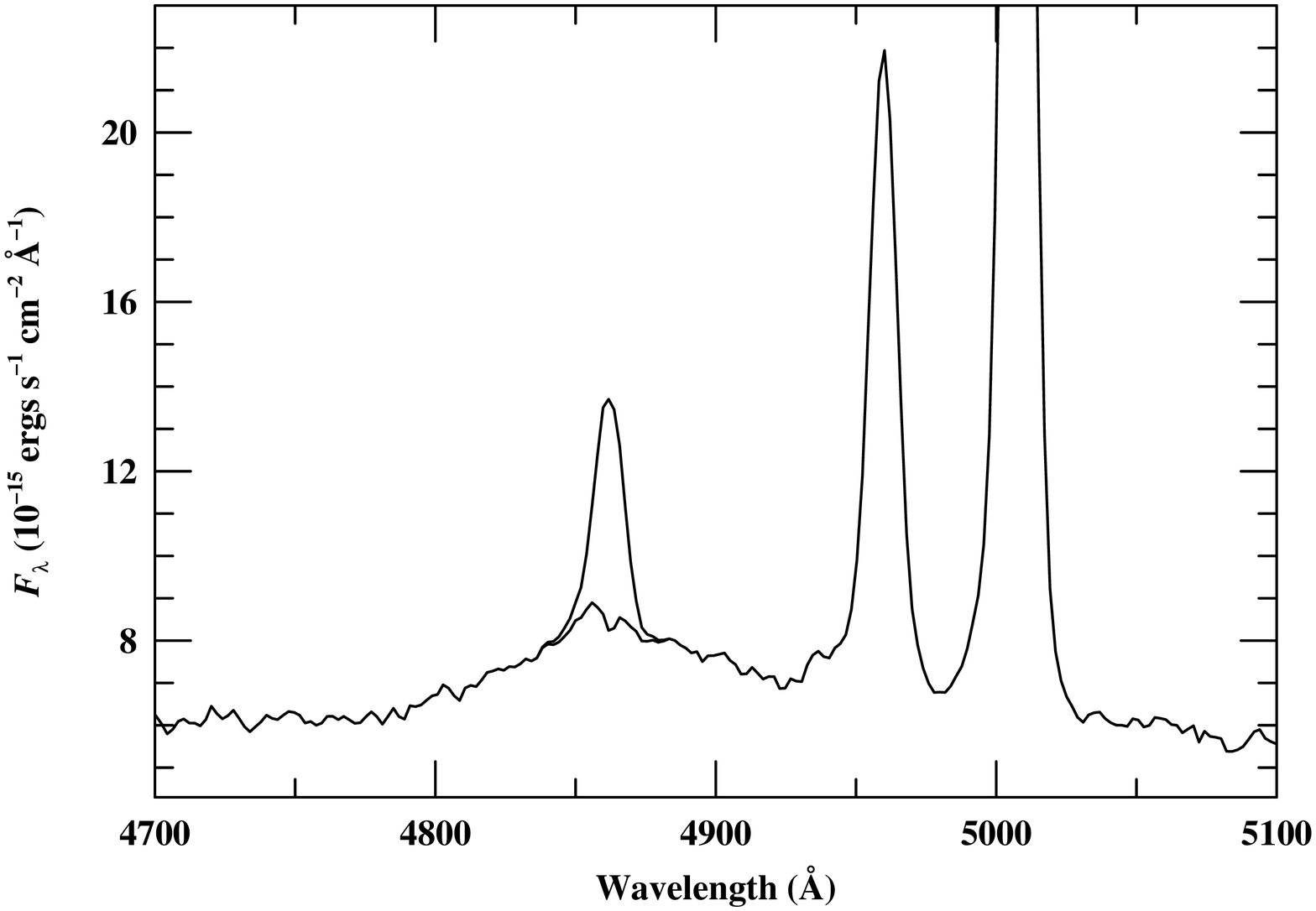}
\caption{The \Hbeta\ spectral region of NGC 5548 in a low state
on JD2448726. The effect of attempting to remove the constant
narrow-line component of \Hbeta\ is shown. The narrow-line
\Hbeta\ component was removed by using the 
\o5007\ line as a template, scaled in flux by a factor of
0.12.}
\end{figure}

\clearpage
\begin{figure}
\epsscale{0.8}
\plotone{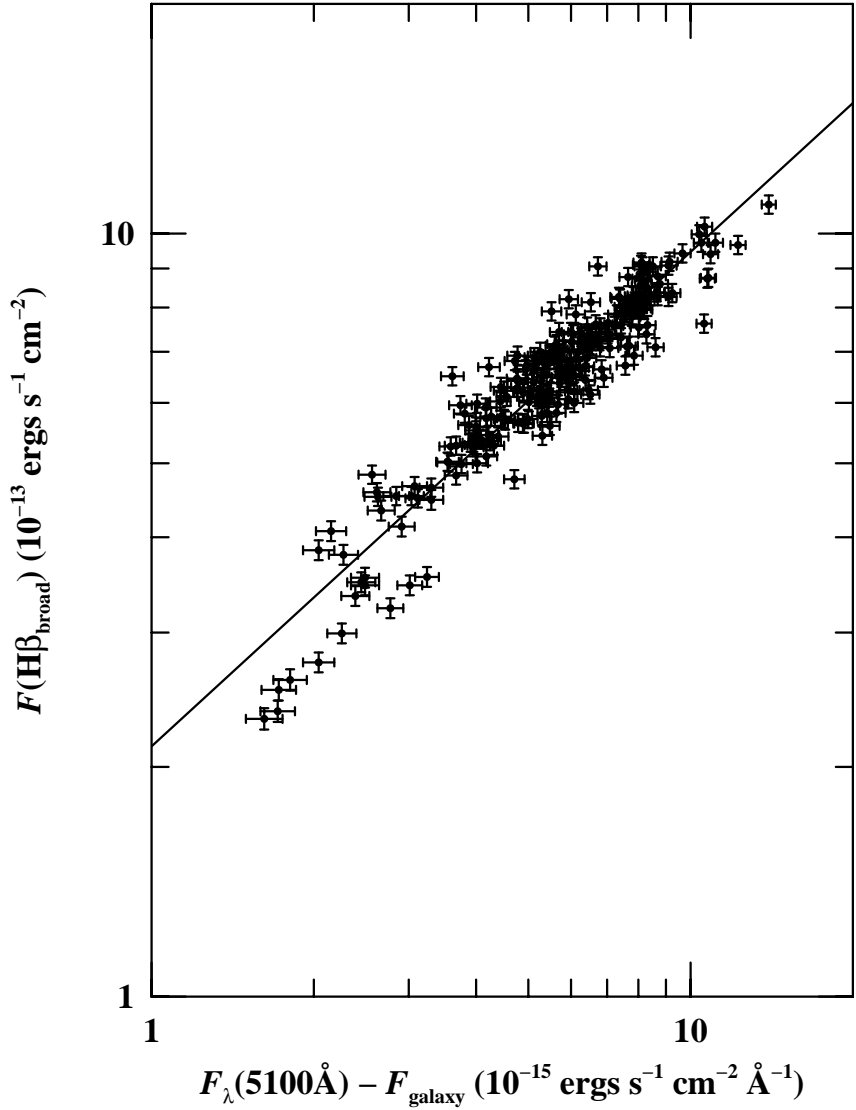}
\caption{Broad \Hbeta\ flux as a function of
optical non-stellar continuum flux for NGC 5548.
The broad-line \Hbeta\ flux is obtained by subtracting
an \Hbeta\ narrow-line contribution of 
$6.7 \times 10^{-14}$\,\lineunits\ from the total line measurement.
The non-stellar continuum flux is obtained by subtracting
a starlight contribution of 
$3.37 \times 10^{-15}$\,\contunits\ from the optical continuum
measurements. The solid line is the best fit of eq.\
(\ref{eq:powerlaw}) to these data, with slope $\alpha = 0.648$
(Table 2).}
\end{figure}

\clearpage

%
%
\begin{deluxetable}{lcccc}
\tablewidth{0pt}
\tablecaption{Fits to the 
Optical Continuum-\Hbeta\ Relationship}
\tablehead{
\colhead{Fit} &
\colhead{$F(\Hbeta_{\rm narrow})$\tablenotemark{a}} &
\colhead{$F_{\rm galaxy}\tablenotemark{b}$} &
\colhead{$A$} &
\colhead{$\alpha$}
\\
\colhead{(1)} &
\colhead{(2)} & 
\colhead{(3)} &
\colhead{(4)} &
\colhead{(5)}
\\
}
\startdata
Non-linear least squares &
$1.86 \pm 0.28$  &
$4.85 \pm 0.08$  & 
$2.53 \pm 0.17$ &
$0.526 \pm 0.022$ \\
(eq.\ \ref{eq:baldwin})
\\
Power-law fit &
$0.67$\,\tablenotemark{c} &
$3.37$\,\tablenotemark{c}  &
$2.18 \pm 0.09$ &
$0.648 \pm 0.022$ \\
(eq.\ \ref{eq:powerlaw})
\enddata
\tablenotetext{a}{\raggedright Units of $10^{-13}$\,\lineunits.}
\tablenotetext{b}{\raggedright Units of $10^{-15}$\,\contunits.}
\tablenotetext{c}{\raggedright Parameter held fixed.}
\end{deluxetable}


\clearpage

\begin{references}
\reference{}Akritas, M.G., \& Bershady, M.A. 1996, ApJ, 470, 706
\reference{}Baldwin, J. 1977, ApJ, 214, 679 
\reference{}Blandford, R.D., \& McKee, C.F. 1982, ApJ, 255, 419
\reference{}Dietrich, M., Hamann, F., Shields, J.C., Constantin, A.,
Vestergaard, M., Chaffee, F., Foltz, C.B., \& Junkkarinen, V.T. 2002,
ApJ, 581, 912
\reference{}Kinney, A.L., Rivolo, A.R., \& Koratkar, A.P. 1990,
ApJ, 357, 338
\reference{}Korista, K., Baldwin, J., Ferland, G., \& Verner, D. 1997, 
ApJS, 108, 401
\reference{}Korista, K.T., \& Goad, M.R. 2001, ApJ, 553, 695
\reference{}Osmer, P.S., Porter, A.C., \& Green, R.F. 1994,
ApJ, 436, 678
\reference{}Peterson, B.\,M., et al. 2002, ApJ, 581, 197
\reference{}Pogge, R. W., \& Peterson, B. M. 1992, AJ, 103, 1084
\reference{}Press, W., Flannery, B., Teukolsky, S., \& Vetterling, W. 
1989, 
in Numerical Recipes (Fortran) (Cambridge: Cambridge University Press), 
p.\ 508
\reference{}Romanishin, W., et al.\ 1995, ApJ, 455, 516
\end{references}
\end{document}